\documentclass[12pt,a4paper]{article}

\usepackage[utf8]{inputenc}
\usepackage{amsmath}
\usepackage{amsfonts}
\usepackage{amssymb}
\usepackage{mathtools}
\usepackage{mathrsfs}
\usepackage{multirow}
\usepackage{graphicx}
\usepackage{subfig}
\usepackage{xcolor}
\usepackage{cite}
\usepackage{hyperref}
\hypersetup{
    colorlinks=true,
    linkcolor=magenta,
    urlcolor=blue,
    citecolor=blue
}
\usepackage{orcidlink}
\usepackage{comment}
\usepackage{footmisc}

\usepackage{parskip} 
\usepackage{geometry}
\newcommand{\refEq}[1]{Eq.\,\eqref{#1}} 
\newcommand{\refEqs}[1]{Eqs.\,\eqref{#1}} 

\newcommand{\abs}[1]{|#1|} 
\newcommand{\ABS}[1]{\left|#1\right|} 
\newcommand{\re}[1]{\text{Re}\left(#1\right)}
\newcommand{\im}[1]{\text{Im}\left(#1\right)}
\newcommand{\Hc}{\text{H.c.}}
\newcommand{\TR}[1]{\text{Tr}\left\{#1\right\}}

\newcommand{\Yuk}[2]{Y_{#1}^{\rm (#2)}}
\newcommand{\Yukc}[2]{Y_{#1}^{\rm (#2)\ast}}
\newcommand{\Yukd}[2]{Y_{#1}^{\rm (#2)\dagger}}
\newcommand{\Yukt}[2]{Y_{#1}^{{\rm (#2)}T}}
\newcommand{\Yukp}[2]{Y_{#1}^{\rm (#2)\prime}}
\newcommand{\Yukpt}[2]{Y_{#1}^{{\rm (#2)\prime}T}}
\newcommand{\yuk}[2]{y_{#1}^{\rm (#2)}}
\newcommand{\YD}[1]{\Yuk{#1}{d}}
\newcommand{\YDc}[1]{\Yukc{#1}{d}}

\newcommand{\yD}[1]{\yuk{#1}{d}}
\newcommand{\YU}[1]{\Yuk{#1}{u}}
\newcommand{\YUc}[1]{\Yukc{#1}{u}}

\newcommand{\yU}[1]{\yuk{#1}{u}}
\newcommand{\YL}[1]{\Yuk{#1}{\ell}}

\newcommand{\YLd}[1]{\Yukd{#1}{\ell}}
\newcommand{\yL}[1]{\yuk{#1}{\ell}}
\newcommand{\YN}[1]{\Yuk{#1}{\nu}}

\newcommand{\YNd}[1]{\Yukd{#1}{\nu}}
\newcommand{\yN}[1]{\yuk{#1}{\nu}}

\newcommand{\baseSD}{\Phi}
\newcommand{\SD}[1]{\baseSD_{#1}^{\phantom{\dagger}}}
\newcommand{\SDd}[1]{\baseSD_{#1}^\dagger}
\newcommand{\SDc}[1]{\baseSD_{#1}^{\phantom{\dagger}\!\!\!\ast}}
\newcommand{\SDC}[1]{\tilde\baseSD_{#1}^{\phantom{\dagger}}}
\newcommand{\baseHSD}{H}
\newcommand{\HSD}[1]{\baseHSD_{#1}^{\phantom{\dagger}}}

\newcommand{\HSDC}[1]{\tilde\baseHSD_{#1}^{\phantom{\dagger}}}
\newcommand{\nsH}{{H}^0}
\newcommand{\nsR}{{R}^0}
\newcommand{\nsI}{{I}^0}
\newcommand{\nh}{{\rm h}}

\newcommand{\nH}{{\rm H}}
\newcommand{\nA}{{\rm A}}
\newcommand{\cH}{{\rm H}^\pm}
\newcommand{\cHm}{{\rm H}^-}
\newcommand{\cHp}{{\rm H}^+}
\newcommand{\mh}{m_{\nh}}

\newcommand{\mH}{m_{\nH}}
\newcommand{\mA}{m_{\nA}}
\newcommand{\mcH}{m_{\cH}}
\newcommand{\wQL}{Q_L^0}\newcommand{\wQLb}{\overline{Q_L^0}}
\newcommand{\wLLb}{\overline{L_L^0}}
\newcommand{\wUR}{u_R^0}
\newcommand{\wDR}{d_R^0}
\newcommand{\wNR}{\nu_R^0}
\newcommand{\wLR}{\ell_R^0}
\newcommand{\wNuL}{\nu_L^0}

\newcommand{\Cb}{c_\beta}
\newcommand{\Sb}{s_\beta}
\newcommand{\Tb}{t_\beta}

\newcommand{\VEV}[1]{\langle #1 \rangle}
\newcommand{\vev}[1]{v_{#1}}
\newcommand{\vevPh}[1]{\theta_{#1}}
\newcommand{\ROTmat}{\mathcal R}\newcommand{\ROTmatT}{\ROTmat^T}
\newcommand{\ROT}[1]{\ROTmat_{#1}}
\newcommand{\CKM}{V}
\newcommand{\V}[1]{{\CKM_{#1}^{\phantom{\ast}}}}
\newcommand{\Vc}[1]{{\CKM_{#1}^\ast}}
\newcommand{\PMNS}{U}
\newcommand{\U}[1]{{\PMNS_{#1}^{\phantom{\ast}}}}
\newcommand{\Uc}[1]{{\PMNS_{#1}^\ast}}
\newcommand{\baseMatO}{O}
\newcommand{\MatO}[1]{\baseMatO_{#1}^{\phantom{T}}}\newcommand{\MatOt}[1]{\baseMatO_{#1}^{T}}
\newcommand{\MatOL}[1]{\MatO{#1_L}}\newcommand{\MatOLt}[1]{\MatOt{#1_L}}
\newcommand{\MatOR}[1]{\MatO{#1_R}}
\newcommand{\OLUt}{\MatOLt{u}}
\newcommand{\OLD}{\MatOL{d}}

\newcommand{\baseMatU}{U}
\newcommand{\MatU}[1]{\baseMatU_{#1}^{\phantom{\dagger}}}\newcommand{\MatUd}[1]{\baseMatU_{#1}^{\dagger}}
\newcommand{\MatUL}[1]{\MatU{#1_L}}\newcommand{\MatULd}[1]{\MatUd{#1_L}}
\newcommand{\MatUR}[1]{\MatU{#1_R}}\newcommand{\MatURd}[1]{\MatUd{#1_R}}
\newcommand{\ULU}{\MatUL{u}}\newcommand{\ULUd}{\MatULd{u}}\newcommand{\URU}{\MatUR{u}}
\newcommand{\ULD}{\MatUL{d}}\newcommand{\ULDd}{\MatULd{d}}\newcommand{\URD}{\MatUR{d}}
\newcommand{\ULNd}{\MatULd{\nu}}\newcommand{\URN}{\MatUR{\nu}}
\newcommand{\ULL}{\MatUL{\ell}}\newcommand{\ULLd}{\MatULd{\ell}}\newcommand{\URL}{\MatUR{\ell}}
\newcommand{\baseMatM}{M}
\newcommand{\baseMatN}{N}
\newcommand{\wMatM}[1]{\baseMatM_{#1}^{0\phantom{\dagger}}}\newcommand{\wMatMd}[1]{\baseMatM_{#1}^{0\dagger}}
\newcommand{\wMatMc}[1]{\baseMatM_{#1}^{0\phantom{\dagger}\!\!\!\ast}}
\newcommand{\MatM}[1]{\baseMatM_{#1}^{\phantom{\dagger}}}\newcommand{\MatMd}[1]{\baseMatM_{#1}^{\dagger}}

\newcommand{\wMatN}[1]{\baseMatN_{#1}^{0\phantom{\dagger}}}

\newcommand{\MatN}[1]{\baseMatN_{#1}^{\phantom{\dagger}}}

\newcommand{\wMU}{\wMatM{u}}
\newcommand{\wNU}{\wMatN{u}}
\newcommand{\wMD}{\wMatM{d}}
\newcommand{\wND}{\wMatN{d}}
\newcommand{\wMN}{\wMatM{\nu}}\newcommand{\wMNd}{\wMatMd{\nu}}
\newcommand{\wMNc}{\wMatMc{\nu}}
\newcommand{\wNN}{\wMatN{\nu}}
\newcommand{\wML}{\wMatM{\ell}}
\newcommand{\wNL}{\wMatN{\ell}}
\newcommand{\mNU}{\MatN{u}}
\newcommand{\mND}{\MatN{d}}
\newcommand{\mMN}{\MatM{\nu}}\newcommand{\mMNd}{\MatMd{\nu}}
\newcommand{\mNN}{\MatN{\nu}}
\newcommand{\mML}{\MatM{\ell}}\newcommand{\mMLd}{\MatMd{\ell}}
\newcommand{\mNL}{\MatN{\ell}}

\graphicspath{{Figs/}}
\pdfoutput=1

\begin{document}

\hfill\begin{minipage}[r]{0.3\textwidth}\begin{flushright} IFIC/23-22 \end{flushright} \end{minipage}

\vspace{0.50cm}
\begin{center}
{\large{\textbf{
Spontaneous CP violation and $\mu-\tau$ symmetry in two-Higgs-doublet models with flavour conservation}}}

\renewcommand*{\thefootnote}{\fnsymbol{footnote}}
\vspace{0.50cm}
João M. Alves \orcidlink{0000-0002-9955-530X},$^{a,}$\footnote{\href{mailto:j.magalhaes.alves@tecnico.ulisboa.pt}{\texttt{j.magalhaes.alves@tecnico.ulisboa.pt}}}
Francisco J. Botella \orcidlink{0000-0003-2235-2536},$^{a,}$\footnote{\href{mailto:Francisco.J.Botella@uv.es}{\texttt{Francisco.J.Botella@uv.es}}}
Carlos Miró \orcidlink{0000-0003-0336-9025},$^{a,}$\footnote{\href{mailto:Carlos.Miro@uv.es}{\texttt{Carlos.Miro@uv.es}}}
Miguel Nebot \orcidlink{0000-0001-9292-7855} $^{a,}$\footnote{\href{mailto:Miguel.Nebot@uv.es}{\texttt{Miguel.Nebot@uv.es}}}
\renewcommand*{\thefootnote}{\arabic{footnote}}
\setcounter{footnote}{0}

\vspace{0.50cm}
\textit{$^a$Departament de Física Teòrica and Instituto de Física Corpuscular (IFIC),\\ Universitat de València -- CSIC, E-46100 Valencia, Spain.}\\
\end{center}

\vspace{0.50cm}
\begin{abstract}
In multi-Higgs-doublet models, requiring simultaneously that (i) CP violation only arises spontaneously, (ii) tree level scalar flavour changing couplings are absent and (iii) the fermion mixing matrix is CP violating, can only be achieved in a very specific manner. A general approach with new clarifying insights on the question is presented. Considering the quark sector, that peculiar possibility is not viable on phenomenological grounds. We show that, considering the lepton sector, it is highly interesting and leads to viable models with $\mu-\tau$ symmetric PMNS matrices. Phenomenological implications of the models, both for Dirac and Majorana (in a type I seesaw scenario) neutrinos, are analysed.
\end{abstract}

\newpage
\section{Introduction\label{SEC:Intro}}
An appealing motivation for the introduction of several Higgs doublets in $SU(2)_L\otimes U(1)_Y$ models is the possibility to have spontaneous CP violation (SCPV) \cite{Lee:1973iz,Lee:1974jb}. One can indeed construct viable scenarios in which the sole origin of CP violation in the CKM matrix is the vacuum \cite{Nebot:2018nqn}. On the other hand, a significant source of concern in multi-Higgs models is the possibility of tree level scalar flavour changing neutral couplings (SFCNC), severely constrained by experiment. 
In the context of Natural Flavour Conservation (NFC) \emph{à la} Glashow and Weinberg \cite{Glashow:1976nt}, reference \cite{Branco:1979pv} addressed the implications of spontaneous CP violation; it was argued that even if CP is broken by the vacuum, in case one requires that SFCNC are absent, the CKM mixing matrix is nevertheless CP conserving for any number of quark families (notice that, although \cite{Branco:1979pv} considered NFC, this argument only involved the absence of SFCNC). Later on, it was pointed out in \cite{Ecker:1987md} and \cite{Gronau:1987xz} that the previous general argument can be evaded, leading to CP violating CKM matrices with very specific properties together with flavour conservation (FC), i.e. absence of SFCNC; this exception, however, is outside the NFC paradigm, where quarks of a given charge receive their mass through the couplings of precisely one neutral Higgs boson. 
However, as discussed in \cite{Ecker:1987md}, this very specific form of the CKM matrix was not phenomenologically viable. In this work, we address the question in the context of the lepton sector and present a simple model where the PMNS mixing matrix, as a consequence of the simultaneous requirement of flavour conservation and CP violation originated by the vacuum, has $\mu-\tau$ symmetry \cite{Harrison:2002et}. It is to be stressed that no additional symmetry is involved in order to enforce a PMNS mixing with that property. The paper is organised as follows. In section \ref{SEC:Generalities} we set our notation discussing general aspects of two-Higgs-doublet models (2HDMs). In section \ref{SEC:SCPV:SFCNC}, we revisit the arguments in \cite{Branco:1979pv}, \cite{Ecker:1987md} and \cite{Gronau:1987xz} concerning FC and SCPV. A general approach to the question is addressed in section \ref{SEC:Gen}. Finally, in section \ref{SEC:LeptonModel} we present our minimal model with Dirac neutrinos, together with a straightforward extension to a type I seesaw scenario, and discuss some relevant phenomenological consequences. Additional aspects are addressed in the appendices: appendix \ref{APP:Scalar} is a concise summary of the most relevant points concerning the scalar sector of a CP conserving 2HDM with SCPV; appendix \ref{APP:RGE} is devoted to an analysis of the stability of the model (and extensions of it) under renormalization group evolution of the Yukawa matrices; and appendix \ref{APP:O3} recalls the basics of the diagonalization of $O(3,\mathbb{R})$ matrices.

\section{Flavour conservation in 2HDMs\label{SEC:Generalities}}
The Yukawa couplings of 3 generations of quarks and 2 Higgs doublets have the following form
\begin{equation}\label{eq:LYuk:00}
-\mathscr L_{\rm Yq}= \wQLb\left(\YU{1}\SDC{1}+\YU{2}\SDC{2}\right)\wUR+\wQLb\left(\YD{1}\SD{1}+\YD{2}\SD{2}\right)\wDR+\Hc,
\end{equation}
where $\wQL$, $\wUR$, $\wDR$ are 3-vectors in generation space (left-handed doublets with subscript $L$ and right-handed singlets with subscript $R$), $\YU{j}$ and $\YD{j}$ are $3\times 3$ Yukawa coupling matrices, and $\SD{j}$ are the scalar (Higgs) doublets, with $\SDC{j}=i\sigma_2\SDc{j}$ ($j=1,2$).
Electroweak symmetry is spontaneously broken by the vacuum expectation values
\begin{equation}\label{eq:vevs}
\VEV{\SD{j}}=\frac{\vev{j}e^{i\vevPh{j}}}{\sqrt 2}\begin{pmatrix}0\\ 1\end{pmatrix},\quad \vev{j},\vevPh{j}\in\mathbb{R}.
\end{equation}
It is helpful to consider the Higgs basis $\{\HSD{1},\HSD{2}\}$ \cite{Donoghue:1978cj,Georgi:1978ri,Botella:1994cs}
\begin{equation}
 \begin{pmatrix} \HSD{1} \\ \HSD{2}\end{pmatrix}=
 \begin{pmatrix} \Cb & \Sb\\ -\Sb & \Cb \end{pmatrix}
 \begin{pmatrix} e^{-i\vevPh{1}}\SD{1}\\ e^{-i\vevPh{2}}\SD{2}\end{pmatrix},
\end{equation}
with $\vev{}^2=\vev{1}^2+\vev{2}^2=(\sqrt 2 G_F)^{-1}\simeq (246\ \text{GeV})^2$, $\Cb=\cos\beta=\vev{1}/\vev{}$, $\Sb=\sin\beta=\vev{2}/\vev{}$ and $\Tb=\tan\beta$, such that $\VEV{\HSD{1}}=\frac{\vev{}}{\sqrt 2}\left(\begin{smallmatrix}0\\ 1\end{smallmatrix}\right)$, $\VEV{\HSD{2}}=\left(\begin{smallmatrix}0\\ 0\end{smallmatrix}\right)$. In this basis, one has
\begin{equation}\label{eq:LYuk:01}
 -\frac{\vev{}}{\sqrt 2}\mathscr L_{\rm{Yq}}=\wQLb\left(\wMU\HSDC{1}+\wNU\HSDC{2}\right)\wUR+\wQLb\left(\wMD\HSD{1}+\wND\HSD{2}\right)\wDR+\Hc,
\end{equation}
with the mass matrices
\begin{equation}\label{eq:MassMatrices:00}
 \wMU=\frac{\vev{}}{\sqrt 2}e^{-i\vevPh{1}}\left(\Cb\YU{1}+\Sb e^{-i\vevPh{}}\YU{2}\right),\quad 
 \wMD=\frac{\vev{}}{\sqrt 2}e^{i\vevPh{1}}\left(\Cb\YD{1}+\Sb e^{i\vevPh{}}\YD{2}\right),
\end{equation}
and the additional Yukawa couplings
\begin{equation}\label{eq:NMatrices:00}
\wNU=\frac{\vev{}}{\sqrt 2}e^{-i\vevPh{1}}\left(-\Sb\YU{1}+\Cb e^{-i\vevPh{}}\YU{2}\right),\quad 
\wND=\frac{\vev{}}{\sqrt 2}e^{i\vevPh{1}}\left(-\Sb\YD{1}+\Cb e^{i\vevPh{}}\YD{2}\right),
\end{equation}
where $\vevPh{}\equiv\vevPh{2}-\vevPh{1}$. Bidiagonalization of the mass matrices in \refEq{eq:MassMatrices:00} proceeds as follows: since $\wMatM{f}\wMatMd{f}$ and $\wMatMd{f}\wMatM{f}$ ($f=u,d$) are hermitian and positive definite with the same set of eigenvalues, there are unitary matrices $\MatUL{f}$ and $\MatUR{f}$ such that
\begin{equation}
\MatULd{f}\wMatM{f}\wMatMd{f}\MatUL{f} =\text{diag}(m_{f_1}^2,m_{f_2}^2,m_{f_3}^2),\quad \MatURd{f}\wMatMd{f}\wMatM{f}\MatUR{f}=\text{diag}(m_{f_1}^2,m_{f_2}^2,m_{f_3}^2),
\end{equation}
with $m_{f_j}$ real positive, and
\begin{equation}
\wMatM{f}\mapsto\MatM{f}=\MatULd{f}\wMatM{f}\MatUR{f}=\text{diag}(m_{f_1},m_{f_2},m_{f_3}).
\end{equation}
The matrices $\MatUL{f}$ and $\MatUR{f}$ are the unitary transformations of fields into the mass bases, namely $f_R^0=\MatUR{f}f_R$ and $f_L^0=\MatUL{f}f_L$, giving rise to a CKM matrix of the form $V=\ULUd\ULD$. Analogously, one has $\wMatN{f}\mapsto\MatN{f}=\MatULd{f}\wMatN{f}\MatUR{f}$ for the Yukawa couplings in \refEq{eq:NMatrices:00}. The off-diagonal elements of the matrices $\mNU$ and $\mND$ encode SFCNC. Therefore, there is FC if the matrices $\wNU$ and $\wND$ are bidiagonalized with the same $\ULU$, $\URU$, $\ULD$, $\URD$:
\begin{equation}
 \text{FC}\ \Leftrightarrow\ \left\{\begin{aligned} & \ULUd\wNU\URU=\mNU=\text{diag}(n_{u1},n_{u2},n_{u3}),\\ & \ULDd\wND\URD=\mND=\text{diag}(n_{d1},n_{d2},n_{d3}),\end{aligned}\right.
\end{equation}
where, in principle, $n_{fj}\in\mathbb{C}$, $f=u,d$, $j=1,2,3$.

If one imposes invariance of $\mathscr L_{\rm{Yq}}$ under standard CP transformations\footnote{We do not consider here non-standard CP transformations.} \cite{Branco:1999fs}, the Yukawa matrices are real $\Yuk{j}{f}=\Yukc{j}{f}$. Similarly, CP invariance requires real coefficients in the scalar potential. Finally, if $\vevPh{}\neq 0,\pi,$ one may have SCPV \cite{Branco:1980sz}.

\section{SCPV and SFCNC\label{SEC:SCPV:SFCNC}}
Considering a CP invariant Lagrangian, that is real Yukawa matrices in \refEq{eq:LYuk:00}, and SCPV, the following ``CP conserving mixing'' argument was presented in \cite{Branco:1979pv}:

\emph{
FC means that the matrices $\wMatM{q}$ and $\wMatN{q}$ ($q=u,d$) are simultaneously bidiagonalized, which is equivalent to $\Yuk{1}{q}$ and $\Yuk{2}{q}$ being bidiagonalized at the same time. Since $\Yuk{1}{q}$ and $\Yuk{2}{q}$ are real, the bidiagonalization is achieved with real orthogonal matrices, $\MatOLt{q}\Yuk{j}{q}\MatOR{q}=\text{\rm diag}(\yuk{j1}{q},\yuk{j2}{q},\yuk{j3}{q})$, $\yuk{jk}{q}\in\mathbb{R}$, implying that $\MatM{q}=\MatOLt{q}\wMatM{q}\MatOR{q}$ and $\MatN{q}=\MatOLt{q}\wMatN{q}\MatOR{q}$ are diagonal. Then, the CKM matrix is $\CKM=R_U\,\OLUt\OLD\,R_D$ with $R_U,R_D$ diagonal rephasing matrices: the CKM matrix is thus essentially real, not CP violating.
}

Convincing as it is, the previous argument can be evaded \cite{Ecker:1987md,Gronau:1987xz}: even if the matrices $\Yuk{1}{q}$, $\Yuk{2}{q}$ are real, they can have complex eigenvalues and in that case they are not necessarily bidiagonalized simultaneously with real orthogonal matrices. In \cite{Ecker:1987md}, a model with 2 Higgs doublets, 4 quark generations and the following Yukawa matrices, was presented:
\begin{equation}\label{eq:4gen:ex:00}
  \YU{j}=O^T\,\begin{pmatrix}\yU{j1}&0&0&0\\ 0&\yU{j2}&0&0\\ 0&0&a_j&b_j\\ 0&0&-b_j&a_j\end{pmatrix},\qquad \YD{j}=\text{diag}(\yD{j1},\yD{j2},\yD{j3},\yD{j4}),
\end{equation}
where $O$ is a real orthogonal matrix and $\YUc{j}=\YU{j}$, $\YDc{j}=\YD{j}$ (the Yukawa Lagrangian $\mathscr L_{\rm Yq}$ for quarks in \refEq{eq:LYuk:00} is CP invariant). The crucial ingredient in \refEq{eq:4gen:ex:00} are the $2\times 2$ blocks in $\YU{j}$, namely
\begin{equation}\label{eq:block2x2}
    B_j=\begin{pmatrix}a_j & b_j\\ -b_j & a_j \end{pmatrix}.
\end{equation}
They obey (no sum over $j$) $B_jB_j^T=B_j^TB_j=(a_j^2+b_j^2)\mathbf{1}_2$ (henceforward, $\mathbf{1}_n$ is the $n\times n$ identity matrix), so that $B_jB_j^T$ and $B_j^TB_j$ have two \emph{degenerate} real eigenvalues $a_j^2+b_j^2$ while $B_j$ has two complex conjugate eigenvalues $a_j\pm ib_j$:
\begin{equation}\label{eq:block2x2:diag:0}
 \mathcal U_{0}^\dagger B_j \mathcal U_{0}= \begin{pmatrix}a_j+ib_j&0\\ 0&a_j-ib_j\end{pmatrix},\quad \text{with}\quad \mathcal U_{0} =\frac{1}{\sqrt 2}\begin{pmatrix}1 & 1\\ i&-i\end{pmatrix}.
\end{equation}
This implies that $\YU{1}$, $\YU{2}$ are simultaneously diagonalized \emph{unitarily}:\footnote{There is still some freedom in $\mathcal U_0$, since a rephasing $\mathcal U_0\mapsto \mathcal U_0\times\text{diag}(e^{i\alpha},e^{i\beta})$ does not change \refEq{eq:block2x2:diag:0}.}
\begin{equation}\label{eq:block2x2:YUdiag:0}
 \ULUd\YU{j}\URU=\text{diag}(\yU{j1},\yU{j2},a_j+ib_j,a_j-ib_j),\quad 
 \ULU=O^T\,\mathcal U_{[34]}\,,\ \URU=\mathcal U_{[34]}\,,
\end{equation}
with 
\begin{equation}
 \mathcal U_{[34]}=\begin{pmatrix}1&0&0&0\\ 0&1&0&0\\ 0&0&\frac{1}{\sqrt 2}& \frac{1}{\sqrt 2}\\ 0&0&\frac{i}{\sqrt 2}&\frac{-i}{\sqrt 2}\end{pmatrix}.
\end{equation}

On the other hand, it is straightforward to check that the simultaneous real orthogonal bidiagonalization of $\YU{1}$ and $\YU{2}$ \emph{fails}. In fact, one can bidiagonalize either $B_1$ or $B_2$ with real orthogonal matrices, but not both simultaneously if $B_1$ and $B_2$ are not proportional --and that is the case if the resulting quark masses are non-degenerate--.

The diagonalization of the mass matrices involves additional (diagonal) rephasings in order to have real mass terms, omitted in the resulting CKM matrix, which is
\begin{equation}\label{eq:exCKM:00}
\CKM =\mathcal U_{[34]}^\dagger\,O\,.
\end{equation}
From \refEq{eq:exCKM:00} it follows that $\V{3j}=\Vc{4j}=(O_{3j}-iO_{4j})/\sqrt{2}$, and thus the rephasing invariant relation 
\begin{equation}
|\V{3j}|=|\V{4j}|,\quad j=1,2,3,4.
\end{equation}
This relation is not phenomenologically viable in a 4 quark generation model; a 3 quark generation implementation of the same idea is not feasible for the same essential reason: experimentally, two CKM elements in the same row/column cannot have the same modulus. However, if instead of the CKM matrix one considers the PMNS lepton mixing matrix, this possibility is open: before addressing it in section \ref{SEC:LeptonModel}, the question is analysed in more generality in the following section.

\section{Evading the ``CP conserving mixing'' argument: a general analysis\label{SEC:Gen}}
As discussed in \cite{Gronau:1987xz}, a necessary ingredient to evade the ``CP conserving mixing'' argument is the presence of complex conjugate eigenvalues in the Yukawa matrices. In a fermion sector (f) with CP invariance at the Lagrangian level, this implies that the matrices $\Yuk{j}{f}\Yukt{j}{f}$ and $\Yukt{j}{f}\Yuk{j}{f}$ have real degenerate eigenvalues, and thus the simultaneous real orthogonal bidiagonalization involved in the ``CP conserving mixing'' argument fails; nevertheless, a simultaneous unitary diagonalization of the Yukawa matrices which involves complex eigenvectors could exist and lead to a CP violating mixing matrix.

We address the question aiming to a more transparent understanding of the conclusions reached in \cite{Gronau:1987xz}, clarifying in the process how the simultaneous unitary diagonalization of the Yukawa matrices arises. We circumscribe our discussion to 3 fermion generations. Since the key ingredient is the degeneracy of the eigenvalues of $\Yuk{j}{f}\Yukt{j}{f}$ and $\Yukt{j}{f}\Yuk{j}{f}$, we discuss separately the case of two degenerate eigenvalues and the case of all three eigenvalues degenerate.

\underline{Two degenerate eigenvalues}\\
Let us consider first that $\Yuk{j}{f}\Yukt{j}{f}$ has eigenvalues $\{\mu_{j1}^2,\mu_{j2}^2,\mu_{j2}^2\}$, with $\mu_{j1},\mu_{j2}\in\mathbb{R}$ and $\mu_{j1}^2\neq\mu_{j2}^2$. Since $\Yuk{j}{f}\Yukt{j}{f}$ is real and symmetric, there exists a $3\times 3$ real orthogonal matrix $O_L$ such that 
\begin{equation}\label{eq:2degenerate:00}
 O_L^T\Yuk{j}{f}\Yukt{j}{f}O_L=\text{diag}(\mu_{j1}^2,\mu_{j2}^2,\mu_{j2}^2).
\end{equation}
This does not fix $O_L$ completely, since \refEq{eq:2degenerate:00} is invariant under
\begin{equation}
 O_L\mapsto O_L \begin{pmatrix}1 & \begin{matrix}0&0\end{matrix}\\ \begin{matrix}0\\ 0\end{matrix} & O_2\end{pmatrix},\quad O_2\in O(2,\mathbb{R}).
\end{equation}
If one writes explicitly
\begin{equation}
 O_L^T\Yuk{j}{f}=\begin{pmatrix}y_j&u_{j1}&u_{j2}\\ w_{j1}& x_{j11} & x_{j12}\\ w_{j2}& x_{j21}&x_{j22}\end{pmatrix}=\begin{pmatrix}y_j&\vec u_j^T\\ \vec w_j& X_j\end{pmatrix},
\end{equation}
one can introduce $O_R\in O(3,\mathbb{R})$ such that\footnote{The point is that one can choose $O_R$ so that its second and third columns, seen as a vectors, are orthogonal to $(y_j,\vec u_j^{T})$.} 
\begin{equation}
 O_L^T\Yuk{j}{f}O_R=\Yukp{j}{f}=\begin{pmatrix}y_j^\prime&\begin{matrix}0& 0\end{matrix}\\\vec w_j^\prime & X_j^\prime\end{pmatrix}
\end{equation}
and thus
\begin{equation}\label{eq:2degenerate:01}
 O_L^T\Yuk{j}{f}\Yukt{j}{f}O_L=\Yukp{j}{f}\Yukpt{j}{f}=\begin{pmatrix}y_j^{\prime 2}&y_j^\prime\vec w_j^{\prime T}\\ y_j^\prime\vec w_j^{\prime} & \vec w_j^{\prime}\vec w_j^{\prime T}+X_j^\prime X_j^{\prime T}\end{pmatrix}.
\end{equation}
Comparing \refEq{eq:2degenerate:00} with \refEq{eq:2degenerate:01}, and assuming $\mu_{j1},\mu_{j2}\neq 0$, one obtains $\vec w_j^{\prime T}=(0, 0)$ and, most importantly, $X_j^\prime X_j^{\prime T}=\mu_{j2}^2\mathbf{1}_2$. The latter condition implies that $(\mu_{j2})^{-1}X_j^\prime$ is a real orthogonal $2\times 2$ matrix. The most general $\Yuk{j}{f}$ such that $\Yuk{j}{f}\Yukt{j}{f}$ has two degenerate eigenvalues is thus of the form
\begin{equation}
\Yuk{j}{f}\propto O_L\begin{pmatrix}y_j & \begin{matrix}0&0\end{matrix}\\ \begin{matrix}0\\ 0\end{matrix} & O_2\end{pmatrix}O_R^{T},
\end{equation}
with $O_L,O_R\in O(3,\mathbb{R})$ and $O_2\in O(2,\mathbb{R})$. The appearance of an orthogonal submatrix is most important. On the one hand, those having $\det O_2=+1$, that is $SO(2,\mathbb{R})$ matrices, are of the form\footnote{One can also address $SO(2,\mathbb{R})$ matrices following the systematic approach of appendix \ref{APP:O3}, which yields the same results.}
\begin{equation}
 O_2\in SO(2,\mathbb{R})\ \Leftrightarrow\ O_2=\begin{pmatrix}\cos \alpha & \sin\alpha\\ -\sin\alpha&\cos\alpha\end{pmatrix},\ \alpha\in [0;2\pi[.
\end{equation}
They have complex conjugate eigenvalues $e^{i\alpha}$ and $e^{-i\alpha}$, with corresponding complex conjugate eigenvectors $\vec v_+=\frac{1}{\sqrt 2}\left(\begin{smallmatrix}1\\ i\end{smallmatrix}\right)$ and $\vec v_-=(\vec v_+)^\ast$. Since $\vec v_+$, $\vec v_-$ are orthonormal, the following unitary diagonalization exists
\begin{equation}\label{eq:SO2R:diag}
 \mathcal U_{0}^\dagger\begin{pmatrix}\cos \alpha & \sin\alpha\\ -\sin\alpha&\cos\alpha\end{pmatrix}\mathcal U_{0}=\text{diag}(e^{i\alpha},e^{-i\alpha}),\ \mathcal U_{0}=\frac{1}{\sqrt{2}}\begin{pmatrix}1 & 1\\ i& -i\end{pmatrix},
\end{equation}
which is the central ingredient of the counterexample presented in \cite{Ecker:1987md} to the ``CP conserving mixing'' argument discussed in section \ref{SEC:SCPV:SFCNC}: it is sufficient to realize that, introducing $\lambda_j=\sqrt{a_j^2+b_j^2}$, $\cos\varphi_j=\frac{a_j}{\lambda_j}$, $\sin\varphi_j=\frac{b_j}{\lambda_j}$, then $a_j\pm ib_j=\lambda_j e^{\pm i\varphi_j}$ and the $B_j$ matrix in \refEq{eq:block2x2} is simply
\begin{equation}\label{eq:Bj}
B_j=\lambda_j\begin{pmatrix}\cos\varphi_j&\sin\varphi_j\\ -\sin\varphi_j & \cos\varphi_j\end{pmatrix},\quad \text{that is }\lambda_j^{-1}B_j\in SO(2,\mathbb{R}).
\end{equation}
 Notice in addition that $\mathcal U_{0}$ in \refEq{eq:SO2R:diag} does not depend on $\alpha$: all $SO(2,\mathbb{R})$ matrices are simultaneously diagonalizable following \refEq{eq:SO2R:diag}. On the other hand, if the orthogonal submatrix of interest satisfies $\det O_2=-1$ instead, it has the form
\begin{equation}
 O_2\in O(2,\mathbb{R}),\ \det O_2=-1\ \Leftrightarrow\ O_2=\begin{pmatrix}\cos \alpha & \sin\alpha\\ \sin\alpha&-\cos\alpha\end{pmatrix},\ \alpha\in [0;2\pi[,
\end{equation}
and presents real eigenvalues $\pm 1$: although the degeneracy in $\Yuk{j}{f}\Yukt{j}{f}$ is still present, there is no simultaneous irreducibly complex unitary diagonalization such as \refEq{eq:SO2R:diag}.

\underline{Three degenerate eigenvalues}\\
Having stressed the role of orthogonal matrices, the case of $\Yuk{j}{f}\Yukt{j}{f}$ and $\Yukt{j}{f}\Yuk{j}{f}$ having all eigenvalues degenerate becomes almost straightforward: one necessarily has $\Yuk{j}{f}\Yukt{j}{f}=\mu_j^2\mathbf{1}_3$ ($\mu_j\neq 0\in\mathbb{R}$), that is $\mu_j^{-1}\Yuk{j}{f}\in O(3,\mathbb{R})$. $O(3,\mathbb{R})$ matrices have two complex conjugate eigenvalues with corresponding complex conjugate eigenvectors, and a real $\pm 1$ eigenvalue corresponding to a real eigenvector. We emphasize two important points (for the sake of clarity, a short reminder of the unitary diagonalization of $O(3,\mathbb{R})$ matrices is provided in appendix \ref{APP:O3}): (i) the eigenvectors are orthonormal and thus a unitary diagonalization of $\Yuk{j}{f}$ exists, (ii) the eigenvectors do not depend on the eigenvalues and thus there are different $O(3,\mathbb{R})$ matrices that can be unitarily diagonalized simultaneously.

One can then construct counterexamples to the ``CP conserving mixing'' argument extending the case analysed in \cite{Ecker:1987md} in a transparent manner via $\Yuk{j}{f}=\mu_j O_j$, $\mu_j\in\mathbb{R}$, $O_j\in O(3,\mathbb{R})$, with all $O_j$ simultaneously diagonalizable unitarily. In \cite{Ecker:1987md,Gronau:1987xz} there is no mention of the intimate connection between degeneracies, complex eigenvalues, a simultaneous unitary diagonalization, and the role of orthogonal matrices: the previous discussion makes it clear (without invalidating, of course, the results in these references).

Concerning the resulting mixing matrices, the crucial common point in the previous two cases ($\Yuk{j}{f}\Yukt{j}{f}$ with two or three degenerate eigenvalues) is the following. Consider the unitary matrix $\mathcal U$ that diagonalizes simultaneously the $\Yuk{j}{f}$ for all $j$; its columns, as vectors, are $\{\vec r,\vec v_+,\vec v_-\}$, which are orthonormal and, crucially, $\vec v_+=(\vec v_-)^\ast$ with both $\vec v_\pm$ non-real, while on the contrary $\vec r$ is real. If the mixing matrix $\mathcal V$ is $\mathcal V=\mathcal U^\dagger O_3$ where $O_3\in O(3,\mathbb{R})$ with real orthonormal column vectors $\vec o_k$, $k=1,2,3$, then $\mathcal V_{2k}=(\vec v_+)^\ast\cdot \vec o_k$ and $\mathcal V_{3k}=(\vec v_-)^\ast\cdot \vec o_k=\vec v_+\cdot\vec o_k=\mathcal V_{2k}^\ast$. It follows that $|\mathcal V_{2k}|=|\mathcal V_{3k}|$ as in section \ref{SEC:SCPV:SFCNC} (starting with a different ordering of the columns of $\mathcal U$, the equality of moduli of $\mathcal V$ elements would apply to the correspondingly permuted rows). If the $\Yuk{j}{f}$ are such that one has instead $\mathcal V=O_3^T\mathcal U$, the equality of moduli is in terms of columns rather than in terms of rows. One can even consider a mixing matrix $\mathcal V=\mathcal U^\dagger \mathcal U^\prime$ with both $\mathcal U$ and $\mathcal U^\prime$ complex unitary matrices with orthonormal columns $\mathcal U\to\{\vec r,\vec v_+,\vec v_-\}$ and $\mathcal U^\prime\to\{\vec r^\prime,\vec v_+^\prime,\vec v_-^\prime\}$, where $(\vec v_{\pm}^{(\prime)})^\ast=\vec v_{\mp}^{(\prime)}$. In this case, rather than equality of moduli in two full different rows/columns, one has $\mathcal V_{31}=\mathcal V_{21}^\ast=(\vec v_-)^\ast\cdot\vec r^\prime$ and $\mathcal V_{13}=\mathcal V_{12}^\ast=\vec r\cdot \vec v_{-}^{(\prime)}$: such a possibility is not phenomenologically viable neither for CKM, nor for PMNS; besides the equality of moduli, the resulting mixing matrix is CP conserving \cite{Gronau:1987xz} (one can readily check that the imaginary part of a rephasing invariant quartet vanishes). It is thus clear how, in general, the basic ingredient underlying the particular example of section \ref{SEC:SCPV:SFCNC} can yield a mixing matrix with the same kind of specific property: equality of moduli in pairs of rows/columns. As mentioned, this kind of property is ruled out in the quark sector, but not in the lepton sector, to which we now turn our attention.

\section{Lepton models\label{SEC:LeptonModel}}
This kind of scenario with FC and a spontaneous origin of CP violation giving an irreducibly complex mixing matrix is not viable in the quark sector. Consequently, we do not discuss the quark sector further and only mention that one would need, in order to have SCPV and a complex CKM, to ensure that SFCNC are under control, as done for example in \cite{Nebot:2018nqn}. Nevertheless, the idea can still work in the lepton sector. In subsection \ref{sSEC:Dirac} we present a simple implementation with Dirac neutrinos; then, in subsection \ref{sSEC:Majorana}, we extend it to a seesaw scenario with Majorana neutrinos, and finally, in subsection \ref{sSEC:Pheno} we discuss some relevant phenomenological consequences concerning neutrinos and charged leptons interactions.

\subsection{Dirac neutrinos\label{sSEC:Dirac}}
For Dirac neutrinos we introduce 3 right-handed neutrino fields $\wNR$, which are singlets under the gauge group, giving rise to the following leptonic Yukawa sector:
\begin{equation}\label{eq:LYukLep:00}
 -\mathscr L_{\rm Y\ell}=\wLLb\left(\YN{1}\SDC{1}+\YN{2}\SDC{2}\right)\wNR+\wLLb\left(\YL{1}\SD{1}+\YL{2}\SD{2}\right)\wLR+\Hc,
\end{equation}
where the Yukawa matrices are given by
\begin{equation}\label{eq:YukLepMat:00}
\YN{j}=\text{diag}(\yN{j1},\yN{j2},\yN{j3}), \quad \YL{j} =O^T\,\lambda_j\begin{pmatrix}\yL{j1}/\lambda_j & 0 & 0\\ 0 & \cos\varphi_j & \sin\varphi_j\\ 0 & -\sin\varphi_j & \cos\varphi_j\end{pmatrix}, 
\end{equation}
with $\lambda_j$, $\varphi_j$, $\yL{j1}$, $\yN{jk}$ real and $O$ a real orthogonal $3\times 3$ matrix. Thus, $\mathscr L_{\rm Y\ell}$ is CP invariant. The procedure is analogous to the steps recalled for the quark sector in section \ref{SEC:Generalities}: transformation of the scalars into the Higgs basis, identification of mass matrices and additional Yukawa matrices and diagonalization of mass matrices; with the first step,
\begin{equation}\label{eq:LYukLep:01}
 -\frac{\vev{}}{\sqrt{2}}\mathscr L_{\rm Y\ell}=\wLLb\left(\wMN\HSDC{1}+\wNN\HSDC{2}\right)\wNR+\wLLb\left(\wML\HSD{1}+\wNL\HSD{2}\right)\wLR+\Hc
\end{equation}
Electroweak and CP invariance are spontaneously broken by the vacuum in \refEq{eq:vevs}. Then, the mass matrices read
\begin{equation}\label{eq:MassMatricesLep:00}
 \wMN=\frac{\vev{}}{\sqrt 2}e^{-i\vevPh{1}}\left(\Cb\YN{1}+\Sb e^{-i\vevPh{}}\YN{2}\right),\quad 
 \wML=\frac{\vev{}}{\sqrt 2}e^{i\vevPh{1}}\left(\Cb\YL{1}+\Sb e^{i\vevPh{}}\YL{2}\right),
\end{equation}
while the additional Yukawa couplings are
\begin{equation}\label{eq:NMatricesLep:00}
\wNN=\frac{\vev{}}{\sqrt 2}e^{-i\vevPh{1}}\left(-\Sb\YN{1}+\Cb e^{-i\vevPh{}}\YN{2}\right),\quad 
\wNL=\frac{\vev{}}{\sqrt 2}e^{i\vevPh{1}}\left(-\Sb\YL{1}+\Cb e^{i\vevPh{}}\YL{2}\right).
\end{equation}
The diagonalization of the charged lepton mass matrix $\wML$ is implemented as
\begin{multline}\label{eq:MassMatricesLep:01}
\wML\mapsto \ULLd\,\wML\,\URL=\mML=\text{diag}(m_e,m_\mu,m_\tau)=\\ \frac{\vev{}}{\sqrt 2}\text{diag}\left(\abs{\Cb\yL{11}+\Sb e^{i\vevPh{}}\yL{21}},\abs{\Cb\lambda_{1}e^{i\varphi_1}+\Sb\lambda_2 e^{i\vevPh{}}e^{i\varphi_2}},\abs{\Cb\lambda_{1}e^{-i\varphi_1}+\Sb\lambda_2 e^{i\vevPh{}}e^{-i\varphi_2}}\right),
\end{multline}
where
\begin{equation}\label{eq:LepDiag:00}
 \ULL=O^T\,\mathcal U_{[23]},\quad \URL=\mathcal U_{[23]}R_{\ell_R},\quad 
 \mathcal U_{[23]}=\begin{pmatrix}1&0&0\\ 0&\frac{1}{\sqrt 2}& \frac{1}{\sqrt 2}\\ 0&\frac{i}{\sqrt 2}& \frac{-i}{\sqrt 2}\end{pmatrix},
\end{equation}
and the rephasing $R_{\ell_R}$ to obtain real diagonal elements is\footnote{As usual, \refEq{eq:MassMatricesLep:01} leaves an additional common rephasing freedom $\URL\mapsto\URL R$, $\ULL\mapsto\ULL R$, with $R=e^{i\,\text{diag}(\delta_1,\delta_2,\delta_3)}$.}
\begin{equation}
\begin{aligned}
 &R_{\ell_R}=e^{-i\vevPh{1}}\text{diag}(e^{-i\gamma_1},e^{-i\gamma_2},e^{-i\gamma_3}),\quad && \gamma_1=\arg(\Cb\yL{11}+\Sb e^{i\vevPh{}}\yL{21}),\\ &\gamma_2=\arg(\Cb\lambda_{1}e^{i\varphi_1}+\Sb\lambda_2 e^{i\vevPh{}}e^{i\varphi_2}),\ &&\gamma_3=\arg(\Cb\lambda_{1}e^{-i\varphi_1}+\Sb\lambda_2 e^{i\vevPh{}}e^{-i\varphi_2}).
 \end{aligned}
\end{equation}
Correspondingly, $\wNL\mapsto\mNL=\ULLd\,\wNL\,\URL$. Although we analyse $\mNL$ in subsection \ref{sSEC:Pheno}, the important point to notice meanwhile is that $\mNL$ is diagonal --there is FC as intended-- but not necessarily real, and this has potential implications for CP violation to be addressed later. The neutrino mass matrix $\wMN$, and also $\wNN$, are already diagonal --only a rephasing is needed to have real eigenvalues in $\mMN=\ULNd\wMN\URN$-- and thus the PMNS matrix (up to rephasings) is
\begin{equation}\label{eq:PMNS:00}
 U=\mathcal U_{[23]}^\dagger O\,.
\end{equation}
Then,
\begin{equation}\label{eq:PMNS:01}
 \U{2j}=\Uc{3j}=\frac{1}{\sqrt 2}(O_{2j}-iO_{3j}),
\end{equation}
leading to the rephasing invariant property 
\begin{equation}\label{eq:PMNS:02}
|\U{2j}|=|\U{3j}|\quad j=1,2,3.
\end{equation}
The PMNS matrix in \refEq{eq:PMNS:02} has ``$\mu-\tau$ symmetry'' \cite{Harrison:2002et}. One salient consequence of this ``$\mu-\tau$ symmetry'' is that, in the standard PDG parametrization \cite{Chau:1984fp,ParticleDataGroup:2022pth}, the angle $\theta_{23}=\pi/4$ and the Dirac CP violating phase is $\delta=\pm\pi/2$. Furthermore, with $\U{1j}=O_{1j}$, the parametric freedom in $O$ allows to fix the values of the remaining parameters  in agreement with experiment \cite{Gonzalez-Garcia:2021dve,deSalas:2020pgw}. Some comments are in order.
\begin{itemize}
 \item It is remarkable, and to some extent counterintuitive, that CP violation in PMNS is independent of the specific value of $\vevPh{}\neq 0,\pi$: that is, no matter the specific value of $\theta$ (as long as $\vevPh{}\neq 0,\pi$), the Dirac CP phase in the PMNS matrix is $\delta=\pm\pi/2$.
 \item On general grounds, if two charged leptons were degenerate, the PMNS matrix would be CP conserving: following \refEq{eq:MassMatricesLep:01}, one thus needs, in terms of the parameters of the model,
 \begin{equation}\label{eq:NonDegenerate:00}
 m_\mu\neq m_\tau\ \Leftrightarrow\ \Cb\Sb\lambda_1\lambda_2\sin(\varphi_2-\varphi_1)\sin\vevPh{}\neq 0\,.
\end{equation}
 In addition to having two relatively complex vevs, i.e. $\Cb\Sb\neq 0$ and $\vevPh{}\neq 0,\pi$, non-degeneracy of the charged leptons requires the presence of $2\times 2$ \refEq{eq:Bj}-like blocks in both $\YL{1}$ and $\YL{2}$, and furthermore these blocks have to be not proportional.
 \item Notice the peculiar fact that in the ``basic'' Jarlskog CP-violation sensitive invariant \cite{Jarlskog:1985ht,Bernabeu:1986fc,Botella:1994cs}, non-fulfilment of \refEq{eq:NonDegenerate:00} would yield $\det[\mML\mMLd,\mMN\mMNd]=0$ because of the (squared) mass differences factors. In this model, checking CP violation at the level of weak basis invariants is crucial \cite{Botella:2012ab}.
 \item \refEq{eq:YukLepMat:00} is the only viable implementation of the basic idea to obtain a complex PMNS matrix with SCPV together with flavour conserving $\YN{j}$, $\YL{j}$: relations like \refEq{eq:PMNS:01} for other rows, or instead for columns if one exchanges $\YN{j}\leftrightarrows\YL{j}$, are in conflict with with the current knowledge of the PMNS matrix \cite{Gonzalez-Garcia:2021dve,deSalas:2020pgw}.
  One can also consider couplings like $\YL{j}$ in \refEq{eq:YukLepMat:00} both for neutrinos and charged leptons: in that case, equality of moduli in two rows/columns of the PMNS matrix would not extend to the whole row/column, but that would nevertheless remain not viable (and would lead to a CP conserving PMNS matrix, contrary to the intended goal).
 \item It is to be noticed that the diagonalization of the neutrino mass matrix has had almost no role in the obtention of the $\mu-\tau$ symmetric PMNS matrix (we have indeed chosen diagonal $\YN{1}$ and $\YN{2}$ to start with).
 \item It is to be stressed that the $\mu-\tau$ symmetry of the PMNS matrix does not derive from an additional fundamental symmetry of the model; as a consequence, as discussed in appendix \ref{APP:RGE}, the minimal implementation of the basic idea --to combine SCPV with FC and CP violating mixings--, is not stable under one loop renormalization group evolution (one loop stability can be nevertheless obtained within non-minimal implementations).
\end{itemize}
We have thus achieved the main goal of this work: \refEq{eq:YukLepMat:00}, together with SCPV, gives a model without tree level SFCNC where the PMNS matrix is not only CP violating, it is phenomenologically viable and has $\mu-\tau$ symmetry. In appendix \ref{APP:RGE} we discuss the stability of this scenario under renormalization group evolution (RGE).

\subsection{Majorana neutrinos -- type I seesaw\label{sSEC:Majorana}}
Attending to the fact commented above that the diagonalization of the neutrino mass matrix has a secondary role in the previous scenario with Dirac neutrinos, it is natural and simple to extend this scenario with the inclusion of Majorana mass terms for the right-handed neutrinos (and thus incorporate, as usual, a rationale for the smallness of neutrino masses). Aiming for simplicity, we introduce diagonal Majorana mass terms
\begin{equation}\label{eq:MajoranaMassTerm:00}
 \mathscr L_{\rm \nu,Maj}=-\frac{1}{2}\left[\overline{(\wNR)^c} M_R \wNR+\overline{\wNR}M_R(\wNR)^c\right],\quad (\wNR)^c=C\overline{\wNR}^T,
\end{equation}
with $M_R=\text{diag}(M_{R1},M_{R2},M_{R3})$, $M_{Rj}\in\mathbb{R}$.
The neutrino mass terms are then
\begin{equation}\label{eq:MajoranaMassTerm:01}
 \mathscr L_{\rm \nu,Mass}= -\frac{1}{2}\begin{pmatrix}\overline{(\wNuL)^c} & \overline{\wNR}\end{pmatrix}\begin{pmatrix} 0 & \wMNc\\ \wMNd & M_R\end{pmatrix}\begin{pmatrix}\wNuL\\ (\wNR)^c\end{pmatrix} + \Hc,\quad \mathscr M=\begin{pmatrix} 0 & \wMNc\\ \wMNd & M_R\end{pmatrix}.
\end{equation}
With $\wMN$ and $M_R$ diagonal, the diagonalization of $\mathscr M$ is reduced to three textbook diagonalizations of $2\times 2$ seesaw blocks of the form
\begin{equation}
 \begin{pmatrix} 0 & \mu_j\\ \mu_j & M_{Rj}\end{pmatrix},\quad \mu_j = e^{i\vevPh{1}}\frac{\vev{}}{\sqrt 2}(\Cb\yN{1j}+\Sb e^{i\vevPh{}}\yN{2j}),\ \mu_j\in\mathbb{C},\ \abs{\mu_j}\ll M_{Rj},
\end{equation}
that is
\begin{equation}\label{eq:MajoranaDiagon:00}
 \mathcal U^T\,\mathscr M\,\mathcal U=\begin{pmatrix}m_{\rm light}& 0\\ 0& m_{\rm heavy}\end{pmatrix},\quad \mathcal U=\begin{pmatrix}C & S\\ -S^\ast & C\end{pmatrix}\begin{pmatrix}R_\nu&0\\ 0&1\end{pmatrix},
\end{equation}
with
\begin{equation}
\begin{aligned}
 &C=\text{diag}(\cos\alpha_1,\cos\alpha_2,\cos\alpha_3),\quad S=\text{diag}(e^{i\beta_1}\sin\alpha_1,e^{i\beta_2}\sin\alpha_2,e^{i\beta_3}\sin\alpha_3),\\
 &R_\nu=i\,\text{diag}(e^{i\beta_1},e^{i\beta_2},e^{i\beta_3}),
\end{aligned}
\end{equation}
where
\begin{equation}\label{eq:Majorana:LHmix}
 \tan 2\alpha_j=2\frac{\abs{\mu_j}}{M_{Rj}}\ll 1,\quad \beta_j=-\arg(\mu_j),
\end{equation}
and
\begin{equation}
 [m_{\rm light}]_{jk}=\delta_{jk}\abs{\mu_j}\tan\alpha_j\simeq \delta_{jk}\frac{\abs{\mu_j}^2}{M_{Rj}},\quad [m_{\rm heavy}]_{jk}=\delta_{jk}M_{Rj}\frac{\tan2\alpha_j}{2\tan\alpha_j}\simeq \delta_{jk}M_{Rj}.
\end{equation}
Being $(\nu_{L{\rm m}})_j$ the 6 mass eigenstates, 3 light with $j=1,2,3$ and 3 heavy with $j=4,5,6$, \refEq{eq:MajoranaDiagon:00} corresponds to
\begin{equation}\label{eq:Neutrinos:LRMass}
 \wNuL=\begin{pmatrix}CR_\nu & S\end{pmatrix}\begin{pmatrix}\nu_{L{\rm m}}\end{pmatrix},\quad \wNR=\begin{pmatrix}-SR_\nu^\ast & C\end{pmatrix}\begin{pmatrix}(\nu_{L{\rm m}})^c\end{pmatrix}.
\end{equation}
Note that, as usual, the light neutrinos are mostly $\wNuL$ and the heavy neutrinos are mostly $\wNR$.

The resulting $3\times 6$ PMNS is
\begin{equation}\label{eq:PMNS:Majorana}
 \PMNS=\ULLd\,\begin{pmatrix}CR_\nu & S\end{pmatrix},
\end{equation}
with $\ULLd$ in \refEq{eq:LepDiag:00}. In \refEq{eq:PMNS:Majorana}, deviations from $3\times 3$ unitarity in the light sector are controlled by the very small $\sin\alpha_j$ factors in $S$, and Majorana phases can be read from $R_\nu$. To conclude with this illustrative scenario with Majorana neutrinos, we stress the important points in common with the previous Dirac neutrino scenario: (i) there is FC in the charged lepton sector, and (ii) the PMNS matrix is not only CP violating, it has $\mu-\tau$ symmetry. SFCNC in the neutrino sector are addressed in the next subsection.

\subsection{Phenomenological implications\label{sSEC:Pheno}}
In this subsection we finally address some phenomenological implications of the previous scenarios. We discuss separately neutrinos and charged leptons.

\subsubsection{Neutrinos\label{ssSEC:Neutrinos}}
In the Dirac neutrino case, both $\wNN$ and $\mNN$ are diagonal and there are no SFCNC. Furthermore, if $\wMN$ in \refEq{eq:MassMatricesLep:00} does not involve large cancellations, one reasonably has $(\mNN)_{jj}\sim m_{\nu_j}$ and thus the flavour conserving Yukawa interactions of neutrinos appear to be safely negligible with a $\frac{m_{\nu_j}}{\vev{}}\sim 10^{-13}$ suppression.

Concerning the Majorana neutrinos of the seesaw scenario, there are some differences with respect to the Dirac neutrino case. First, since the Yukawa interactions arise from $\wLLb\wNN\HSDC{2}\wNR+\Hc$ terms in \refEq{eq:LYukLep:01}, and $\wNuL$ ($\wNR$) are mostly light (heavy) neutrinos, the Yukawa couplings involving only light neutrinos are suppressed by $\sin\alpha_j$ in \refEq{eq:Majorana:LHmix}. In addition, since one has a scenario with three independent pairs with one light and one heavy neutrino each (mixings and couplings between different pairs are absent), there are in fact no SFCNC in the light-light sector. There are, nevertheless, SFCNC involving one light and one heavy neutrino (one non-vanishing coupling for each independent pair): since our concern lies with effects not at the large scale $M_{R_j}$, we do not discuss them further.

As a final comment on the neutrino sector, while the original motivation --obtaining a CP violating PMNS mixing matrix when one requires that (i) the only source of CP violation is the vacuum state, and that (ii) there are no SFCNC-- fully applies in the Dirac neutrino case, this motivation is weakened in the second scenario with Majorana neutrinos, since neutrino masses arise differently (they involve a different diagonalization). However, it is remarkable that the most relevant features of the Dirac neutrino case in subsection \ref{sSEC:Dirac}, that is a $\mu-\tau$ symmetric PMNS and no SFCNC for charged leptons, are preserved in the simple seesaw Majorana scenario of subsection \ref{sSEC:Majorana}.

\subsubsection{Charged leptons\label{ssSEC:ChLeptons}}
The most important sources of potential phenomenological concern are the flavour conserving couplings of the charged leptons in $\mNL$. From \refEqs{eq:MassMatricesLep:00}--\eqref{eq:MassMatricesLep:01}, with $\varphi=\varphi_2-\varphi_1$, we have
\begin{equation}\label{eq:Nl:diag:00}
 \begin{aligned}
  &n_e \equiv (\mNL)_{ee}=\frac{\vev{}^2}{2m_e}\left(\Cb\Sb([\yL{21}]^2-[\yL{11}]^2)+\yL{11}\yL{21}(\Cb^2-\Sb^2)\cos\vevPh{}+i\yL{11}\yL{21}\sin\vevPh{}\right),\\
  &n_\mu \equiv (\mNL)_{\mu\mu}=\frac{\vev{}^2}{2m_\mu}\left(\Cb\Sb(\lambda_2^2-\lambda_1^2)+\lambda_{1}\lambda_{2}(\Cb^2-\Sb^2)\cos(\vevPh{}+\varphi)+i\lambda_{1}\lambda_{2}\sin(\vevPh{}+\varphi)\right), \\
  &n_\tau \equiv (\mNL)_{\tau\tau}=\frac{\vev{}^2}{2m_\tau} \left(\Cb\Sb(\lambda_2^2-\lambda_1^2)+\lambda_{1}\lambda_{2}(\Cb^2-\Sb^2)\cos(\vevPh{}-\varphi)+i\lambda_{1}\lambda_{2}\sin(\vevPh{}-\varphi)\right).
 \end{aligned}
\end{equation}
The charged lepton masses appear explicitely in \refEq{eq:Nl:diag:00} because, from \refEq{eq:MassMatricesLep:01}, we have used
\begin{equation}\label{eq:Ml:diag:00}
 \begin{aligned}
  & \frac{2}{\vev{}^2}m_e^2=\Cb^2[\yL{11}]^2+\Sb^2[\yL{21}]^2+2\Cb\Sb\yL{11}\yL{21}\cos\vevPh{},\\
  & \frac{2}{\vev{}^2}m_\mu^2=\Cb^2\lambda_{1}^2+\Sb^2\lambda_{2}^2+2\Cb\Sb\lambda_{1}\lambda_{2}\cos(\vevPh{}+\varphi),\\
  & \frac{2}{\vev{}^2}m_\tau^2=\Cb^2\lambda_{1}^2+\Sb^2\lambda_{2}^2+2\Cb\Sb\lambda_{1}\lambda_{2}\cos(\vevPh{}-\varphi).
 \end{aligned}
\end{equation}
The couplings of the charged leptons with the physical neutral scalars (see appendix \ref{APP:Scalar} for further details) are
\begin{equation}\label{eq:Sll:couplings:00}
 \mathscr L_{S\bar{\ell}\ell}=-\frac{1}{\vev{}}S_j\,\bar \ell(a_j^\ell+ib_j^\ell\gamma_5)\ell,\quad \ell=e,\mu,\tau,
\end{equation}
with $j=1,2,3,$ corresponding to $\{S_1,S_2,S_3\}=\{\nh,\nH,\nA\}$, and
\begin{equation}\label{eq:Sll:couplings:01}
 \begin{aligned}
& a_j^\ell=\ROT{1j}m_\ell+\ROT{2j}\re{n_\ell}-\ROT{3j}\im{n_\ell},\\
& b_j^\ell=\ROT{2j}\im{n_\ell}+\ROT{3j}\re{n_\ell},
 \end{aligned}
\end{equation}
where $\ROT{jk}$ is the scalar mixing in \refEq{eq:Scalar:R}.

\begin{figure}[h!tb]
\begin{center}
\subfloat[One loop.\label{sfig:eEDM:1loop}]{\includegraphics[width=0.3\textwidth]{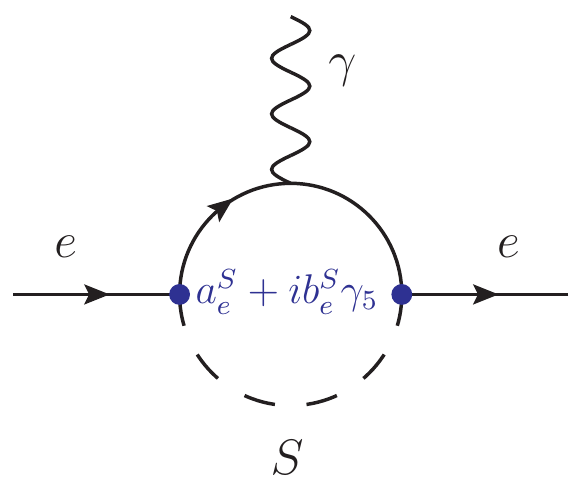}}\quad
\subfloat[Two loops (Barr-Zee).\label{sfig:eEDM:2loop}]{\includegraphics[width=0.3\textwidth]{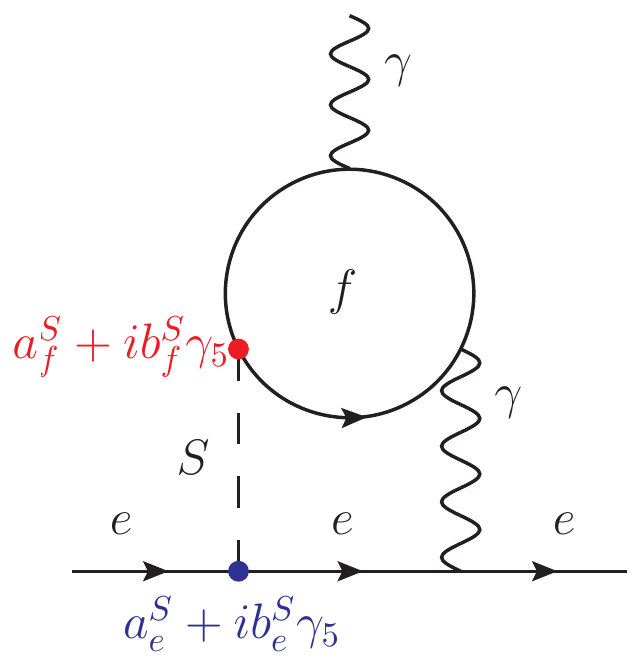}}
\caption{Loop contributions to the eEDM.\label{fig:eEDM}}
\end{center}
\end{figure}

The observables that can be affected and deserve some attention are listed below.
\begin{itemize}
 \item Signal strengths of the 125 GeV SM Higgs-like scalar in $\ell^+\ell^-$ decays.
 \item Similarly, scalar resonances in $pp\to S\to\ell^+\ell^-$ for $q^2\sim m_{S}^2$ (with $q^2$ the invariant mass of the lepton pair); one can also consider production of the charged scalar $\cH$ and decays $\cHp\to\ell^+\nu$, $\cHm\to\ell^-\bar\nu$, but they are controlled by the same couplings and the experimental sensitivity is poorer.
 \item Contributions to precision observables sensitive to loops involving the new scalars and the couplings in \refEq{eq:Nl:diag:00}, such as the $g-2$ of the electron and the muon ($g$ is the gyromagnetic ratio), and the electric dipole moment of the electron (eEDM).
\end{itemize}
One could address them through a full numerical exploration of the parameter space of the model (which in principle has enough parametric freedom to satisfy those constraints), but that is beyond the scope of this work --establishing that the model can be viable-- and would require, in addition, that the quark sector is also specified. Fortunately, references \cite{Botella:2020xzf,Botella:2022rte,Botella:2023tiw} can provide a satisfactory answer. In these works, a general flavour conserving 2HDM is considered in order to address deviations from SM expectations in $g-2$ of both the electron and the muon. In the full numerical analyses presented in \cite{Botella:2020xzf,Botella:2022rte,Botella:2023tiw}, signal strengths of the SM Higgs-like and bounds on resonant production of the new scalars are relevant constraints. These constraints are satisfied despite the fact that the diagonal entries of $\mNL$ required to obtain sizable contributions to $(g-2)_{e,\mu}$ are much larger than the corresponding charged lepton masses (in addition, one important ingredient is that the scalar sector has to be close to the scalar alignment limit). Since here we do not necessarily require such values of the couplings, no problem is to be expected from these observables. The only exception is the eEDM $d_e$, since CP conservation is assumed in \cite{Botella:2020xzf,Botella:2022rte,Botella:2023tiw}. Let us analyse the eEDM in some detail. The experimental constraint is $|d_e|<1.1\times 10^{-29}$ e $\cdot$ cm $=5.6\times 10^{-16}$ e $\cdot$ GeV$^{-1}$. With the couplings in \refEq{eq:Sll:couplings:01}, the contribution of one loop diagrams like Figure \ref{sfig:eEDM:1loop} to $d_e$ is
\begin{equation}\label{eq:eEDM:1loop:00}
 \abs{d_e}=\frac{e}{8\pi^2m_e\vev{}^2}\ABS{\sum_{j=\nh,\nH,\nA}a_j^eb_j^e\ I\left(\frac{m_e^2}{m_{S_j}^2}\right)},
 \quad I\left(\frac{m_e^2}{m_{S_j}^2}\right)\simeq \frac{m_e^2}{m_{S_j}^2}\left(\frac{3}{2}+\ln\left(\frac{m_e^2}{m_{S_j}^2}\right)\right),
\end{equation}
that is 
\begin{equation}\label{eq:eEDM:1loop:01}
 \abs{d_e}\simeq\frac{em_e}{8\pi^2\vev{}^2}\ABS{\sum_{j=\nh,\nH,\nA}\frac{a_j^eb_j^e}{m_{S_j}^2}\,\left(\frac{3}{2}+\ln\left(\frac{m_e^2}{m_{S_j}^2}\right)\right)}.
\end{equation}
As a guidance to how problematic the constraint could be, consider the contribution of a scalar with mass $m_{S}\simeq 500$ GeV and couplings $a_S^e=b_S^e\simeq 10m_e$; with these values,
\begin{equation}\label{eq:eEDM:1loop:02}
 \abs{d_e^{(S)}}=\frac{em_e}{8\pi^2\vev{}^2}\ABS{\frac{a_S^eb_S^e}{m_{S}^2}\,\left(\frac{3}{2}+\ln\left(\frac{m_e^2}{m_{S}^2}\right)\right)}\simeq 2.9\times 10^{-19}\ \text{e}\cdot\text{GeV}^{-1}.
\end{equation}
It is thus clear that, even considering a scalar $S$ not too heavy together with enhancements in both the scalar and pseudoscalar couplings $a_S^e$ and $b_S^e$ with respect to $m_e$, these contributions are not expected to be in conflict with the experimental constraint. One might worry then about Barr-Zee two loop contributions, where the key point is that one of the ``small'' electron Yukawa couplings is replaced by the Yukawa coupling of an internal fermion (which can compensate other loop suppression factors). The contribution of the illustrative diagram in Figure \ref{sfig:eEDM:2loop} to $d_e$ is
\begin{equation}\label{eq:eEDM:2loop:00}
 \abs{d_e^{(S,f)}}=\frac{e\alpha N_c^fQ_f^2}{8\pi^3\vev{}^2m_f}\ABS{b^e_Sa^f_S\,F\left(\frac{m_f^2}{m_S^2}\right)+a^e_Sb^f_S\,G\left(\frac{m_f^2}{m_S^2}\right)},
\end{equation}
\begin{equation}
F(z)=\frac{z}{2}\int_0^1 dx\,\frac{1-2x(1-x)}{x(1-x)-z}\,\ln\left(\frac{x(1-x)}{z}\right)\,,
\end{equation}
\begin{equation}
G(z)=\frac{z}{2}\int_0^1 dx\,\frac{1}{x(1-x)-z}\,\ln\left(\frac{x(1-x)}{z}\right)\,,
\end{equation}
where $N_c^f$ and $Q_f$ are the number of colours and the electric charge of fermion $f$, $a_f^S$ and $b_f^S$ are the scalar and pseudoscalar couplings of $f$ (as in \refEq{eq:Sll:couplings:00}), and $F$ and $G$ are loop functions. With the previous values $m_S\simeq 500$ GeV, $a_S^e=b_S^e\simeq 10m_e$, and an internal top quark $f=t$ with couplings $a_S^t=b_S^t\simeq m_t$, we have $\abs{d_e^{(S,t)}}\sim 3\times 10^{-13}\ \text{e}\cdot\text{GeV}^{-1}$, which is incompatible with the experimental bound. From this rough comparison it is clear that the eEDM is indeed a relevant constraint and these illustrative values (considered simultaneously) are too naive or generic: although not ``automatically granted'', compatibility with experiment can be regained by a combination of heavier scalars, smaller Yukawa couplings of the electron,\footnote{Smaller Yukawa couplings are in fact to be expected since the couplings in \refEq{eq:Sll:couplings:01} involve the interplay between the real and imaginary parts of $(\mNL)_{ee}$ and the scalar mixings $\ROT{jk}$.} smaller Yukawa couplings of the top quark (unless the bottom quark contribution is much enhanced by the couplings, the dominant contribution is the top quark one), and even cancellations from the contributions of the different scalars.
As an example consider the joint contribution of two scalars $S_1$ and $S_2$ with masses $m_{S_1}=m_{S_2}=1$ TeV, couplings to the electron given by \refEq{eq:Sll:couplings:01} with $n_e=\frac{e^{i\pi/4}}{15}m_e$, couplings to the top quark analogous to \refEq{eq:Sll:couplings:01} with $\ell\mapsto t$, $\ROT{3j}\mapsto -\ROT{3j}$ (this change of sign is simply due to the opposite weak isospin of up quarks and charged leptons) and $n_t=\frac{e^{i\pi/4}}{15}m_t$, and finally scalar mixings $\ROT{22}=\ROT{33}=\cos\theta_{23}$, $\ROT{23}=-\ROT{32}=\sin\theta_{23}$ with $\theta_{23}=0.2$. Considering these illustrative values, which have similar size scalar and pseudoscalar couplings both for $e$ and $t$, one has $|d_e^{(S_1,t)}+d_e^{(S_2,t)}|\simeq 3.7\times 10^{-16}\ \text{e}\cdot\text{GeV}^{-1}$, in agreement with the experimental bound.
The main point, to close the discussion, is that the model remains viable but the eEDM plays an important role in shaping the available parameter space.

One might also worry about potentially dangerous contributions to $\mu\to e\gamma$ transitions similar to Figure \ref{fig:eEDM}. At one loop, rather than the diagram \ref{sfig:eEDM:1loop} with a neutral scalar and an electron running in the loop (with the external photon attached to the electron) one has an internal charged scalar and a neutrino (the external photon now attached to the scalar). At two loops, diagram \ref{sfig:eEDM:2loop} is modified replacing the neutral scalars $S$ with the charged scalar $\cH$ and the internal photon with a $W^\mp$. For Dirac neutrinos, the smallness of neutrino masses and couplings $(\mNN)_{jj}\sim m_{\nu_j}$ safely suppress $\mu\to e\gamma$ transitions. For Majorana neutrinos in the scenario of section \ref{sSEC:Majorana}, one might worry about the contributions with internal ``mostly heavy'' $\wNR$. One expects again suppressed contributions from the large heavy neutrino masses in the one loop diagrams, and (less transparently, see \cite{Ilisie:2015tra}) from the heavy neutrino in the initial-to-final fermion line, together with the heavier $W^\mp$ in the internal vector boson line. In any case, although plausible, this does not guarantee automatic agreement with experimental constraints on $\mu\to e\gamma$ transitions, and, as in the case of the electron EDM, they can play a relevant role in shaping the available parameter space in more detailed phenomenological studies.

\section{Conclusions\label{SEC:Conc}}
In the context of multi-Higgs-doublet models where CP invariance holds at the Lagrangian level, the simultaneous requirement of (i) absence of SFCNC and (ii) spontaneous CP violation, is analysed. In \cite{Branco:1979pv}, it was argued that these two conditions necessarily lead to a CP conserving mixing matrix. As shown in \cite{Ecker:1987md}, the Yukawa couplings of a very particular model can reconcile both requirements with a CP violating mixing matrix. Further aspects of the question were addressed in \cite{Gronau:1987xz}. There are two central ingredients. First, the combinations $\Yuk{j}{f}\Yukt{j}{f}$ and $\Yukt{j}{f}\Yuk{j}{f}$ must have (real) \emph{degenerate} eigenvalues. Second, $\Yuk{j}{f}$ must have complex conjugate eigenvalues with corresponding complex conjugate eigenvectors leading to a \emph{simultaneous unitary} diagonalization of all $\Yuk{j}{f}$ ($j=1, 2\ldots$), which guarantees the absence of SFCNC. A general approach to the model presented in \cite{Ecker:1987md} and the analysis presented in \cite{Gronau:1987xz} has been addressed in this work, stressing the important role of orthogonal matrices and their intimate connection with degeneracies, complex eigenvalues and a simultaneous unitary diagonalization of the Yukawa couplings. We present the only viable implementation of this idea in the lepton sector, considering both Dirac and Majorana neutrinos (the latter in a type I seesaw scenario). In all cases, the resulting PMNS mixing matrix, as a consequence of the simultaneous requirement of SFCNC absence and SCPV, is experimentally viable and $\mu-\tau$ symmetric. Phenomenological implications of the model are discussed, with particular emphasis on the role of the electric dipole moment of the electron in shaping the available parameter space.

\section*{Acknowledgments\label{SEC:Ack}}
The authors have received support from \textit{Agencia Estatal de Investigaci\'on}-\textit{Ministerio de Ciencia e Innovaci\'on} (AEI-MICINN, Spain) under grants PID2019-106448GB-C33 and PID2020-113334GB-I00/AEI/10.13039/501100011033 (AEI/FEDER, UE) and from \textit{Generalitat Valenciana} under grant PROMETEO 2019-113 and CIPROM 2022-36.\\ JA is supported by the \textit{GenT Plan} from \textit{Generalitat Valenciana} under project CIDEGENT/2018/014. CM is funded by \textit{Conselleria de Innovación, Universidades, Ciencia y Sociedad Digital} from \textit{Generalitat Valenciana} and by \textit{Fondo Social Europeo} under grants ACIF/2021/284 and CIBEFP/2022/92. MN is supported by the \textit{GenT Plan} from \textit{Generalitat Valenciana} under project CIDEGENT/2019/024.

\appendix
\section{Scalar sector\label{APP:Scalar}}
The CP conserving 2HDM scalar potential reads
\begin{equation}\label{eq:2HDMpotCP:00}
\begin{aligned}
 \mathscr V(\SD{1},\SD{2})=\ 
 &\mu_{11}^2\SDd{1}\SD{1}+\mu_{22}^2\SDd{2}\SD{2}+\mu_{12}^2(\SDd{1}\SD{2}+\SDd{2}\SD{1})+\lambda_1(\SDd{1}\SD{1})^2+\lambda_2(\SDd{2}\SD{2})^2\\
 &+2\lambda_3(\SDd{1}\SD{1})(\SDd{2}\SD{2})+2\lambda_4(\SDd{2}\SD{1})(\SDd{1}\SD{2})+\lambda_5((\SDd{1}\SD{2})^2+(\SDd{2}\SD{1})^2)\\
 &+\lambda_6(\SDd{1}\SD{1})(\SDd{1}\SD{2}+\SDd{2}\SD{1})+\lambda_7(\SDd{2}\SD{2})(\SDd{1}\SD{2}+\SDd{2}\SD{1}),
\end{aligned}
\end{equation}
where $\mu_{ij}^2$, $\lambda_j\in\mathbb{R}$. With $V(\vev{1},\vev{2},\vevPh{})= \mathscr V(\VEV{\SD{1}},\VEV{\SD{2}})$ and $\VEV{\SD{j}}$ in \refEq{eq:vevs}, the stationarity conditions for the vacuum allow to trade\footnote{As discussed in \cite{Nebot:2019qvr}, if one imposes perturbativity constraints on the quartic couplings $\lambda_j$'s, this forces the scalar masses to be bounded from above. The absence of a decoupling regime can be avoided, however, by allowing soft CP violating terms $\im{\mu_{12}^2}\neq 0$ in \refEq{eq:2HDMpotCP:00}.}
\begin{equation}
 \begin{aligned}
 & \mu_{12}^2=-\frac{\vev{}^2}{2}[4\lambda_5\Cb\Sb\cos\vevPh{}+\lambda_6\Cb^2+\lambda_7\Sb^2],\\  
 & \mu_{11}^2=-\vev{}^2[\lambda_1\Cb^2+(\lambda_3+\lambda_4-\lambda_5)\Sb^2+\lambda_6\Cb\Sb\cos\vevPh{}],\\
 & \mu_{22}^2=-\vev{}^2[\lambda_2\Sb^2+(\lambda_3+\lambda_4-\lambda_5)\Cb^2+\lambda_7\Cb\Sb\cos\vevPh{}].
 \end{aligned}
\end{equation}
Expanding the fields around the vacuum
\begin{equation}
\SD{j}=\frac{e^{i\vevPh{j}}}{\sqrt 2}\begin{pmatrix}\sqrt{2}\phi_j^+\\ \vev{j}+\rho_j+i\eta_j\end{pmatrix},\quad
\HSD{1}=\frac{1}{\sqrt{2}}\begin{pmatrix}\sqrt{2}G^+\\ \vev{}+\nsH+iG^0\end{pmatrix},\quad 
\HSD{2}=\frac{1}{\sqrt{2}}\begin{pmatrix}\sqrt{2}\cHp\\ \nsR+i\nsI\end{pmatrix},
\end{equation}
 one can identify the would-be Goldstone bosons $G^0$, $G^\pm$, and read the mass terms for the new neutral scalars (for $\cH$ one can readily obtain $\mcH^2=\vev{}^2(\lambda_5-\lambda_4)$) as
 \begin{equation}
\frac{1}{2} \begin{pmatrix}\nsH & \nsR & \nsI\end{pmatrix}\,\mathcal M_0^2\,\begin{pmatrix}\nsH & \nsR & \nsI\end{pmatrix}^T,
 \end{equation}
with $\mathcal M_0^2$ real and symmetric. The neutral scalars mass matrix $\mathcal M_0^2$ is diagonalized with a real $3\times 3$ orthogonal matrix $\ROTmat$
\begin{equation}\label{eq:Scalar:R}
\ROTmatT\mathcal M_0^2\ROTmat=\text{diag}(\mh^2,\mH^2,\mA^2),\quad \begin{pmatrix}\nh\\ \nH\\ \nA\end{pmatrix}=\ROTmatT\begin{pmatrix}\nsH \\ \nsR \\ \nsI\end{pmatrix},
\end{equation}
which gives the physical neutral scalars $\{\nh,\nH,\nA\}$. 
The important point to recall here is that, since
\begin{equation}
\begin{aligned}
 &[\mathcal M_0^2]_{13}=-\vev{}^2\sin\vevPh{}[4\lambda_5\Sb\Cb\cos\vevPh{}+\lambda_6\Cb^2+\lambda_7\Sb^2],\\
 &[\mathcal M_0^2]_{23}=\vev{}^2\sin\vevPh{}[2\lambda_5(\Cb^2-\Sb^2)\cos\vevPh{}+(\lambda_7-\lambda_6)\Sb\Cb],
 \end{aligned}
\end{equation}
are in principle non-zero, $\ROTmat$ ``mixes'' all neutral scalars and we do not have, as usual in other contexts, scalars $\nh,\nH$ and pseudoscalar $\nA$ (as clearly seen in \refEqs{eq:Sll:couplings:00}-\eqref{eq:Sll:couplings:01}).

\section{Renormalization group evolution\label{APP:RGE}}
As commented in precedence, the $\mu-\tau$ symmetric character of the PMNS matrix does not derive from an imposed additional symmetry. It is thus interesting to analyse the one loop renormalization group evolution of this kind of scenario, paying special attention to two aspects: (i) despite being absent at tree level, how SFCNC may arise at one loop; (ii) is it possible to depart from the minimal implementation of the model (while maintaining the essential ingredients) and still obtain absence of SFCNC at one loop?

The one loop renormalization group evolution (RGE) of the lepton Yukawa couplings is
\begin{multline}\label{eq:RGE:YL:00}
 \mathcal D \YL{a}=c_{\ell}\YL{a}+\sum_{j=1}^{n_d}\left(3\TR{\YL{a}\YLd{j}+\YNd{a}\YN{j}}+\TR{\YL{a}\YLd{j}+\YNd{a}\YN{j}}\right)\YL{j}\\ +\sum_{j=1}^{n_d}\left[-2\YN{j}\YNd{a}\YL{j}+\YL{a}\YLd{j}\YL{j}+\frac{1}{2}\YN{j}\YNd{j}\YL{a}+\frac{1}{2}\YL{j}\YLd{j}\YL{a}\right],
\end{multline}
\begin{multline}\label{eq:RGE:YN:00}
 \mathcal D \YN{a}=c_{\nu}\YN{a}+\sum_{j=1}^{n_d}\left(3\TR{\YL{a}\YLd{j}+\YNd{a}\YN{j}}+\TR{\YL{a}\YLd{j}+\YNd{a}\YN{j}}\right)\YN{j}\\ +\sum_{j=1}^{n_d}\left[-2\YL{j}\YLd{a}\YN{j}+\YN{a}\YNd{j}\YN{j}+\frac{1}{2}\YL{j}\YLd{j}\YN{a}+\frac{1}{2}\YN{j}\YNd{j}\YN{a}\right],
\end{multline}
where $c_{\ell}$ and $c_{\nu}$ are linear combinations of squared $SU(3)_c$, $SU(2)_L$ and $U(1)_Y$ gauge couplings, $\mathcal D\equiv 16\pi^2\frac{d}{d\ln\mu}$ with $\mu$ the energy scale, and $n_d$ is the number of Higgs doublets.

It is rather clear from \refEqs{eq:RGE:YL:00}--\eqref{eq:RGE:YN:00} that in principle SFCNC will arise through one loop RGE. If FC is imposed at some scale $\mu_\mathrm{FC}$, one would thus expect SFCNC at a scale $\mu$ proportional to $\ln(\mu/\mu_\mathrm{FC})$, which might be sufficient to keep them under control. In both \refEqs{eq:RGE:YL:00} and  \eqref{eq:RGE:YN:00}, the terms that can spoil FC are the ones with products of 3 Yukawa coupling matrices within square brackets. Half of these terms are nevertheless harmless. With $\YL{a}$ and $\YN{a}$ in \refEq{eq:YukLepMat:00}, the terms $\YL{a}\YLd{j}\YL{j}$ and $\YL{j}\YLd{j}\YL{a}$ have the same structure of $\YL{a}$ and are simultaneously diagonalized unitarily, while $\YN{a}\YNd{j}\YN{j}$ and $\YN{j}\YNd{j}\YN{a}$, like $\YN{a}$, are diagonal. From the point of view of FC, the problematic terms which give rise to SFCNC through RGE are $\YN{j}\YNd{a}\YL{j}$, $\YN{j}\YNd{j}\YL{a}$, $\YL{j}\YLd{a}\YN{j}$ and $\YL{j}\YLd{j}\YN{a}$. If one thinks of extending the simple model of section \ref{sSEC:Dirac} in order to preserve FC at one loop, it is by no means impossible to consider scenarios where this is achieved, as we discuss for illustration in the following. First, one can observe that if more than two Higgs doublets are considered and if no Higgs doublet couples simultaneously to $\wLR$ and $\wNR$, then all terms $\YN{j}\YNd{a}\YL{j}$ and $\YL{j}\YLd{a}\YN{j}$ vanish. This does not eliminate the terms $\YN{j}\YNd{j}\YL{a}$ and $\YL{j}\YLd{j}\YN{a}$; in fact, since $\YN{j}\YNd{j}$ and $\YL{j}\YLd{j}$ are guaranteed to be non-vanishing, these terms must be present. However, they will not give rise to SFCNC if $\YN{j}\YNd{j}\propto \mathbf{1}_3$ and $\YL{j}\YLd{j}\propto \mathbf{1}_3$. An example scenario with Dirac neutrinos where PMNS is CP violating, there is FC and the only source of CP violation is SCPV, and where SFCNC do not arise through one loop RGE is the following: a model with 5 Higgs doublets with real Yukawa couplings
\begin{equation}
 \YL{1}=\lambda_1\,O_{\ell 1},\quad \YL{2}=\lambda_2\,O_{\ell 2},\quad \YL{j}=\mathbf{0},\ j=3,4,5,
\end{equation}
\begin{equation}
\begin{aligned}
 &\YN{j}=\mathbf{0},\ j=1,2,\\ 
 &\YN{3}=O_\nu^T\text{diag}(y,0,0),\quad\YN{4}=O_\nu^T\text{diag}(0,y,0),\quad\YN{5}=O_\nu^T\text{diag}(0,0,y),
 \end{aligned}
\end{equation}
where $O_{\ell 1},O_{\ell 2},O_\nu\in O(3,\mathbb{R})$, and $O_{\ell 1}$, $O_{\ell 2}$, have the same eigenvectors but different eigenvalues (see appendix \ref{APP:O3}).
The vevs $\VEV{\SD{j}}$ ($j=3,4,5$) need to be chosen such that $y\vev{3}/\sqrt{2}=m_{\nu 1}$, $y\vev{4}/\sqrt{2}=m_{\nu 2}$, $y\vev{5}/\sqrt{2}=m_{\nu 3}$ (that is, these vevs are responsible of the hierarchy/ordering of the neutrino masses). We then have in \refEqs{eq:RGE:YL:00} and \eqref{eq:RGE:YN:00}
\begin{equation}
 \sum_{j=1}^5\YN{j}\YNd{a}\YL{j}=\mathbf{0},\quad \sum_{j=1}^5\YL{j}\YLd{a}\YN{j}=\mathbf{0},
\end{equation}
and
\begin{equation}
\begin{aligned}
 &\sum_{j=1}^5\YNd{j}\YN{j}=\sum_{j=1}^5\YN{j}\YNd{j}=y^2\mathbf{1}_3,\\
 &\sum_{j=1}^5\YLd{j}\YL{j}=\sum_{j=1}^5\YL{j}\YLd{j}=(\lambda_1^2+\lambda_2^2)\mathbf{1}_{3},
\end{aligned}
\end{equation}
which give both $\mathcal D\YL{a}$ and $\YL{a}$, and $\mathcal D\YN{a}$ and $\YN{a}$, simultaneously diagonalizable, and thus no SFCNC arise through one loop RGE.

\section{$O(3,\mathbb{R})$ matrices\label{APP:O3}}
Matrices $O\in SO(3,\mathbb{R})$ are of the form $O=\exp (\alpha A)$ with real $\alpha\in[0;2\pi[$ and $A$ a normalized antisymmetric matrix\footnote{The elements of $A$ are $[A]_{ab}=\epsilon_{abc}\hat n_c$, with $\epsilon_{abc}$ the antisymmetric Levi-Civita tensor.}
\begin{equation}\label{eq:O3:A:00}
 A=\begin{pmatrix} 0&\hat n_3& -\hat n_2\\ -\hat n_3&0&\hat n_1\\ \hat n_2&-\hat n_1&0\end{pmatrix},\quad \hat n_j\in\mathbb{R},\quad 
 \hat n_1^2+\hat n_2^2+\hat n_3^2=1.
\end{equation}
The eigenvalues of A and their corresponding normalized eigenvectors are
\begin{equation}
 \begin{aligned}
   &\lambda_0=0,\ \vec v_{0}^T=(\hat n_1,\hat n_2,\hat n_3)=(\sin\vartheta\cos\varphi,\sin\vartheta\sin\varphi,\cos\vartheta),\\
   &\lambda_{+}=i,\ \vec v_{+}^T=\frac{1}{\sqrt 2}(-\cos\vartheta\cos\varphi+i\sin\varphi,-\cos\vartheta\sin\varphi-i\cos\varphi,\sin\vartheta),\\
   &\lambda_{-}=-i,\ \vec v_{-}=(\vec v_{+})^\ast,
 \end{aligned}
\end{equation}
where, for convenience, a parametrization of the real unit vector $(\hat n_1,\hat n_2,\hat n_3)$ with spherical-like coordinates is used. The set $\{\vec v_0,\vec v_+,\vec v_-\}$ is orthonormal,\footnote{This is guaranteed since $\vec v_j$ ($j=0,\pm$) are eigenvectors of the real symmetric semipositive definite matrix $A^TA$.} and thus we have the unitary diagonalization
\begin{equation}
\mathcal U^\dagger\,A\,\mathcal U=\text{diag}(0,i,-i),\qquad 
 \mathcal U=\begin{pmatrix}\uparrow &\uparrow&\uparrow\\ \vec v_0&\vec v_+&\vec v_-\\ \downarrow&\downarrow&\downarrow\end{pmatrix}.
\end{equation}
For $O=\exp(\alpha A)$ it is immediate that the eigenvalues are $1$, $e^{i\alpha}$ and $e^{-i\alpha}$, the corresponding eigenvectors are $\vec v_0$, $\vec v_+$ and $\vec v_-$ above, and its unitary diagonalization reads
\begin{equation}
\mathcal U^\dagger\,O\,\mathcal U= \text{diag}(1,e^{i\alpha},e^{-i\alpha}).
\end{equation}
Geometrically $O$ represents a rotation in $\mathbb{R}^3$ of angle $\alpha$ around the axis $(\hat n_1,\hat n_2,\hat n_3)$ (in terms of $A$, one has the compact expression $O=1+A\sin\alpha+A^2(1-\cos\alpha)$).

This construction makes clear that the eigenvectors of $O$ do not depend on $\alpha$; this implies, in particular, that $O_1=\exp(\alpha_1 A)$ and $O_2=\exp(\alpha_2 A)$ with $\alpha_1\neq\alpha_2$ and the same $A$ in \refEq{eq:O3:A:00}, can be unitarily diagonalized simultaneously.

Finally, if $O'\in O(3,\mathbb{R})$ with $\det O'=-1$ (i.e. $O'\notin SO(3,\mathbb{R})$), the previous discussion applies straightforwardly by noticing that $-O'\in SO(3,\mathbb{R})$.

\bibliographystyle{JHEP}
\bibliography{revised_FC-SCPV.bib}

\end{document}